\documentclass[3p,times, preprint,
a4paper,sort,compress]{elsarticle}
\usepackage{epsf}
\usepackage{amsmath}
\usepackage{wrapfig}

\def\beq{\begin{equation}}
\def\eeq{\end{equation}}
\def\bea{\begin{eqnarray}}
\def\eea{\end{eqnarray}}
\def\beqa{\begin{equation}\begin{array}{l}}
\def\eeqa{\end{array}\end{equation}}
\def\eqlab#1{\label{eq:#1}}
\def\figlab#1{\label{fig:#1}}

\def\Eqref#1{Eq.~(\ref{eq:#1})}

\def\Figref#1{Fig.~\ref{fig:#1}}

\def\slad{\partial \hspace{-1.7mm} \slash\,}

\def\half{\mbox{\small{$\frac{1}{2}$}}}

\def\thalf{\mbox{\small{$\frac{3}{2}$}}}
\def\quarter{\mbox{\small{$\frac{1}{4}$}}}
\def\third{\mbox{\small{$\frac{1}{3}$}}}

\def\barr{\left(\begin{array}{c}}
\def\earr{\end{array}\right)}
\def\bmat{\left(\begin{array}{cc}}
\def\emat{\end{array}\right)}
\def\al{\alpha}
\def\be{\beta}
\def\ga{\gamma} 
\def\de{\delta} \def\De{\Delta}\def\vDe{\varDelta}
\def\veps{\varepsilon}  \def\eps{\epsilon}

\def\la{\lambda} \def\La{{\Lambda}}

 \def\Si{{\it\Sigma}}
\def\th{\theta}  
  
\def\vfi{\varphi}

\def\dd{{\rm d}}
\def\pa{\partial}

\def\pa{\partial}

\def\nn{\nonumber}

\def\cO{\mathcal{O}}

\def\lag{{\mathcal L}}

\def\mathscr{\mathcal}

\def\re{\mbox{Re}}
\def\im{\mbox{Im}}
\def\3d{3-D}

\def\ol#1{\overline{#1}}

\def\s#1{\setbox0=\hbox{$#1$}  
   \dimen0=\wd0     
   \setbox1=\hbox{/} \dimen1=\wd1  
   \ifdim\dimen0>\dimen1   
      \rlap{\hbox to \dimen0{\hfil/\hfil}} 
      #1     
   \else     
      \rlap{\hbox to \dimen1{\hfil$#1$\hfil}} 
      /      
   \fi}      %

\begin{document}

\hfill Preprint MKPH-T-10-02

\title{A dispersion relation for the pion-mass dependence of hadron properties}
\author{Tim Ledwig}
\author{Vladimir Pascalutsa}
\author{Marc Vanderhaeghen}
\address{Institut f\"ur Kernphysik, Johannes Gutenberg Universit\"at, Mainz D-55099, Germany}


\begin{abstract}
We present a dispersion relation in the pion-mass squared, which
static quantities (nucleon mass, magnetic moment, etc.) obey  under the assumption
of analyticity in the entire complex $m_\pi^2$ plane modulo a cut
at negative  $m_\pi^2$ associated with pion production. The relation is verified here in
 a number of examples of nucleon and $\Delta$-isobar properties
computed in chiral perturbation theory up to order $p^3$. We outline a
method to obtain relations for other mass-dependencies,
and illustrate it on a two-loop example.
\end{abstract}

\maketitle

\section{Introduction}
Present lattice QCD calculations are still limited to larger than physical
values of light quark masses, $m_q > m_{u,d}\simeq 5-10$ MeV, but
the chiral perturbation theory ($\chi$PT) \cite{Weinberg:1978kz, Gasser:1983yg} can, in many cases, be applied to bridge the gap
between the lattice and the real world (see {\it e.g.},~\cite{Bernard:2002yk,Leinweber:2001ui,Pascalutsa:2005ts,Hemmert:2003cb,Wang07,Bernard:2006gx}). 
$\chi$PT can predict
at least some 
of the `non-analytic' dependencies of static quantities (masses, magnetic moments, etc.)
on pion-mass squared, or the quark mass ($m_\pi^2 \sim m_q$). The rest of 
the  contributions contain the {\it a priory} unknown low-energy constants (LECs). 
In this paper we examine the origins of non-analytic dependencies arising in $\chi$PT,
by considering analytic properties of the chiral expansion in the complex
$m_\pi^2$ plane.
\smallskip 

\begin{wrapfigure}{r}{2.2in}
\vspace{-4mm}
\centerline{\epsfclipon   \epsfxsize=4.cm%
  \epsffile{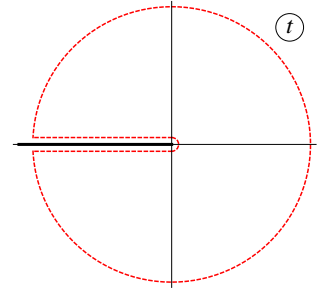} 
}
\caption{The cut and the contour in the complex $t=m_\pi^2$ plane, which go into the
derivation of the dispersion relation in \Eqref{disprel}.}
\figlab{cont1}
\end{wrapfigure}
The basic observation is that  chiral loops exhibit a cut along the {\it negative} $m_\pi^2$
axis.
The cut is associated with pion production which can occur without any excess of energy for
 $m_\pi^2 \leq 0$. Assuming
analyticity in the rest of the $m_\pi^2$-plane (see \Figref{cont1}), one arrives at a dispersion
relation of the type:
\beq
\eqlab{disprel}
 f(m_\pi^2) = -\frac{1}{\pi}\int\limits_{-\infty}^0 \dd t \, \frac{\im\, f(t)}{t-m_\pi^2+i0^{+}}\,,
\eeq
where $f$ is a static quantity, $0^+$ is an infinitesimally small positive number. In what follows, we explicitly verify this type of dispersion relation on 
a few examples of the  nucleon and $\De(1232)$-isobar properties
and discuss its field of application. In particular, we consider
a two-loop example (a sunset graph) for which the absorptive part can relatively easy
be extracted. We conclude by comparing this dispersion relation with
a similar ``mass-dispersion" relation long-known in the literature.

\bigskip

\section{Nucleon mass}

We begin right away by considering the
{\it nucleon} properties as a function of $t=m_\pi^2$. For example, the pion-mass dependence of
the nucleon mass, computed to the $n$th order in the chiral expansion, can be written as:
\beq
M_N =  \sum_{\mathrm{even}\,\, \ell }^n \!\! a_\ell\, t^{\frac{\ell}{2}} + 
 \sum_{\ell }^n \Sigma^{(\ell)}_N(t)\, , 
\eeq
where $a$'s are some linear combinations of LECs,
 $\Sigma_N^{(\ell)}(t)$ is the $\ell$th order nucleon self-energy given
by the graphs of the type shown in \Figref{nuclse}.
\begin{figure}[tb]
\centerline{\epsfclipon   \epsfxsize=11.5cm%
  \epsffile{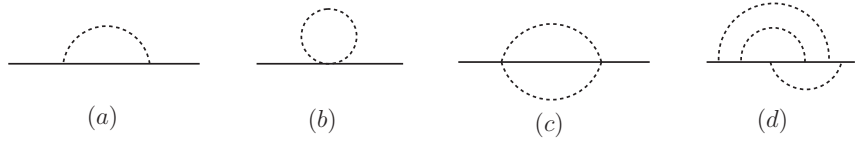} 
}
\caption{Graphs representing chiral-loop corrections to the nucleon mass. Nucleon (pion) propagators are denoted by solid (dashed) lines.}
\figlab{nuclse}
\end{figure}
According to the  power counting rules~\cite{Wein96}, a graph with 
$L$ loops, $N_\pi$ pion and $N_N$ nucleon lines, 
$V_k$ vertices from the Lagrangian of order $k$, 
 contributes at order $p^n$, with $p$ being the generic light
 scale and 
\beq
\eqlab{pc}
n = \sum_k  k V_k + 4L - 2N_\pi -N_N\, .
\eeq
The leading order pion-nucleon Lagrangian is
of order $k=1$, and, to the first order in the pion-field $\pi^a(x)$ (with index $a=1,2,3$), 
is written as \cite{GSS89}:
\beq
\lag^{(1)}_{\pi N} = \bar N(x)\,\bigg( i \slad -\stackrel{\,\circ}{M}_{N}  
+   \frac{\stackrel{\circ}{g}_A}{2\! \stackrel{\circ}{f}_\pi} \,  
(\slad \pi^a) \tau^a \ga_5 \bigg)\, N(x) + \mbox{c.t.} + \cO(\pi^2), 
\eqlab{Nlagran}
\eeq
where $N(x)$ is the isospin-doublet nucleon field, $\tau^a$ are Pauli matrices, 
$\stackrel{\circ}{M}_{N}$, 
$ \stackrel{\circ}{g}_A$, and $\stackrel{\circ}{f}_\pi$ are respectively:
the nucleon mass, axial-coupling and pion-decay constants, in the
chiral limit ($m_\pi\to 0$); ``c.t." stands for counter-term contributions,
which are required for the renormalization of the nucleon mass, field, and so on.

The self-energy receives its leading contribution 
at order $p^3$, which is given by the graph \Figref{nuclse}(a) and 
the following expression:
\begin{subequations}
\bea
\Sigma^{(3)}_N (t) & = &\left. \frac{3 g_A^2}{4f_\pi^2} \,i \! \int \!\frac{\dd^4 k }{(2\pi)^4}
\frac{k \cdot\ga \,\gamma_5 ( p\cdot\gamma - k\cdot\ga+M_{N}) k \cdot\ga \,\gamma_5}{(k^2-t+i0^{+}) [(p-k)^2-M_{N}^2+i0^{+}]}\, \right|_{p\cdot \ga = M_N}
\eqlab{se0}\\
& \stackrel{\mathrm{dim reg}}{=} & 
\!\!-\frac{3 g_A^2}{4f_\pi^2} \frac{M_{N}^3}{(4\pi)^2} \int\nolimits_0^1\! \dd x\, \Big\{
[x^2+ (1-x)\tau] \left(L_\veps+\ln [x^2+ (1-x) \tau- i0^{+}]\right) \nn\\
& & \hskip2.1cm + \, [2x^2-(2+x)\tau]  \left(L_\veps+1+\ln [x^2+ (1-x)\tau- i0^{+}] \right)
- 3L_\veps \Big\},
\eqlab{selfen}
\eea
\end{subequations}
where  $\tau=t/M_N^2$, $L_\veps = -1/\veps -1 + \gamma_E - \ln(4\pi \La/M_{N})$
exhibits the ultraviolet (UV) divergence as $\veps=(4-d)/2 \to 0$, with $d$ being the number of space-time dimensions, $\La$  the scale of dimensional regularization, and $\gamma_{E}\simeq 0.5772$ the Euler's constant. 
Note that for simplicity we assume the physical values for the parameters: 
$M_N\simeq 939$ MeV, $g_A\simeq 1.267$, $f_\pi\simeq 92.4$ MeV;
the difference with the chiral-limit values leads to  higher order effects.

After integration over the Feynman-parameter $x$, this result can be written as: 
\begin{subequations}
\bea
&& \Sigma^{(3)}_N (t) = \frac{3 g_A^2 M_{N}^3}{2(4\pi f_\pi)^2}\Big\{ -L_\veps
+\left(1-L_\veps\right) \frac{t}{M_N^2} \Big\}\,+\,\overline \Sigma^{(3)}_N,
\eqlab{low} \\
\mbox{with} && \overline \Sigma^{(3)}_N (t) = - \frac{3 g_A^2 M_{N}^3}{(4\pi f_\pi)^2}
 \Big( \tau^{3/2} \sqrt{1-\quarter \tau } \, \,\arccos (\half\! \sqrt{\tau}\,) + \quarter \tau^2\,  \ln \tau \Big) .
  \eqlab{renorm}
\eea
\end{subequations}
The term in figure brackets, containing the UV-divergence,
 must be entirely canceled by the counter-term
contribution~\cite{Gegelia:1999gf}, which, to this order, is of the form: $a_0 + a_2\, t,$
where $a$'s contain the ``bare'' values of the LECs. The first term in brackets can be viewed
as a renormalization of the nucleon  mass, while the second as a renormalization
of the $\pi N$ sigma term.
The remaining part, $\ol \Sigma_N$,
is UV-finite and consistent with the power counting in the sense that its size is indeed of 
order $p^3$.

Let us now see whether this contribution obeys the dispersion relation
of the type stated in \Eqref{disprel}. The imaginary part can be easily found from 
\Eqref{selfen} by taking into account that $\ln(-1+i0^{+}) = i\pi$,
\beq
\eqlab{impart}
\mathrm{Im}\, \Sigma_N^{(3)} (t) = 
 \frac{3 g_A^2}{(4\pi f_\pi)^2} \frac{\pi}{2} \left[ - (-t)^{3/2} \left( 1-\frac{t}{4M_N^2} \right)^{1/2}
\! + \,\frac{t^2}{2 M_N}\right] \, \th(-t)\,,
\eeq
where $\th$ is the step function. It is quite obvious that the dispersion integral with
this imaginary part diverges, which is consistent with the fact that the self-energy is
UV divergent.  From \Eqref{low} we have seen that the divergencies appear in the first
two orders of the expansion around $t=0$ and are subsequently absorbed by the counter-terms.
As the result one needs to make two subtractions at point $t=0$ to find
\beq
 \Sigma_N(t)  - 
  \Sigma_N(0) -   \Sigma_N' (0)\, \, t \, =\, 
 -\frac{1}{\pi} \int\limits_{-\infty}^0 \dd t' \,\frac{\mathrm{Im}\, \Sigma_N (t') }{t'-t+i0^{+} } \left(\frac{t}{t'}\right)^2.
\eeq
Now the dispersion integral, with the imaginary part given by \Eqref{impart}, converges,  and moreover, gives the result identical
to the expression in \Eqref{renorm}, that is, 
the renormalized self-energy
contribution. The subtractions have played here the role of the counter-terms. 
We therefore conclude that the  order-$p^3$ self-energy correction to
the nucleon mass obeys the suitably subtracted dispersion relation of the type of
\Eqref{disprel}. We emphasize that the subtractions do not introduce any additional 
uncertainty in the result. The number of subtractions is not arbitrary
but corresponds with the number of counter-terms available at a given order.

\section{Magnetic moment and polarizability}
We next turn to the example of chiral corrections to the nucleon's magnetic moment.
For this we introduce the electromagnetic interaction,  firstly by the minimal substitution
[i.e., $ \pa_\mu N \to \pa_\mu N + \half(1+\tau_3) e A_\mu N$, $\pa_\mu \pi^a \to \pa_\mu \pi^a + \eps^{ab3} e A_\mu \pi^b$, with $e\simeq \sqrt{4\pi/137}\,$]
in the chiral Lagrangian, and secondly by writing out the relevant non-minimal terms: 
\beq
\lag^{(2)}_{\pi N} = -  \frac{e}{4M_N} 
\,\ol N\, \Big( \half(1+\tau_3)\stackrel{\circ}{\kappa}_p+ \half(1-\tau_3)
\stackrel{\circ}{\kappa}_n \Big) \,\ga^{\mu\nu}\, N \,F_{\mu\nu} + \mbox{c.t.},
\eeq
where $\stackrel{\circ}{\kappa}_p$ and $\stackrel{\circ}{\kappa}_n$ are the chiral-limit
values of the proton's and neutron's anomalous magnetic moment (a.m.m.), respectively; 
furthermore $\ga_{\mu\nu} =
\half \ga_{[\mu} \ga_{\nu]}$, 
$F_{\mu\nu} =
\pa_{[\mu} A_{\nu]}$. 
\begin{figure}[tb]
\centerline{\epsfclipon   \epsfxsize=10.5cm%
  \epsffile{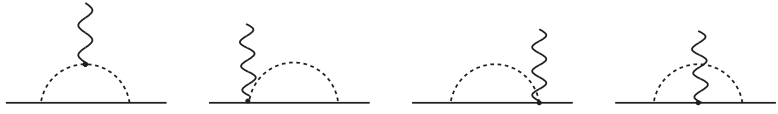} 
}
\caption{Graphs representing order-$p^3$ corrections to the electromagnetic interaction of the nucleon. }
\figlab{emvertex}
\end{figure}
According to the power-counting of \Eqref{pc}, the leading order
chiral correction to the electromagnetic coupling comes at order $p^3$ and is given by the
graphs in \Figref{emvertex}. These graphs give the following contribution to the a.m.m.\ of, respectively, the proton and the neutron:
\begin{subequations}
\bea
\kappa_p^{(p^3\,\mathrm{loop})} (t)&=&
\frac{g_A^2 M_N^2 }{(4\pi f_\pi)^2} \int_0^1 \!\dd x \, \frac{(1-\thalf x)\,x^2}{(1-x)(t/M_N^2) +x^2-i0^+} \, ,\\
\kappa_n^{(p^3\,\mathrm{loop})} (t)&=&
-\frac{g_A^2 M_N^2 }{(4 \pi f_\pi)^2} \int_0^1 \!\dd x \, \frac{x^2}{(1-x) (t/M_N^2)+x^2-i0^+} \,.
 \eea
 \eqlab{ammloop}
 \end{subequations}
For negative $t$, these functions develop an imaginary part:
\begin{subequations}
\bea
\im\, \kappa_p^{(p^3\,\mathrm{loop})} (t)&=&
\frac{g_A^2 M_N^2 }{(4\pi f_\pi)^2}  \frac{\pi  }{2 \la}  \Big(\half \tau + \la \Big)^2
\Big(1-\thalf ( \half \tau +\la ) \Big) \, \th(-t)\, , \\
\im\, \kappa_n^{(p^3\,\mathrm{loop})} (t)&=&
-\frac{g_A^2 M_N^2 }{(4 \pi f_\pi)^2}  \frac{\pi  }{2 \la}  \Big(\half \tau + \la \Big)^2\, \th(-t)\, , 
 \eea
  \end{subequations}
 with $\tau = t/ M_N^2$ and $\la = \sqrt{\quarter \tau^2 -\tau}$. Substituting these expression
 into the dispersion relation \Eqref{disprel}, and performing the integral, we obtain:
 \begin{subequations}
 \bea
 -\frac{1}{\pi}\int\limits_{-\infty}^0 \dd t^\prime\, \frac{\im \,
  \kappa_p^{(p^3\,\mathrm{loop})}(t^\prime)}{t^\prime - t }&=&
\frac{g_A^2 M_N^2 }{(4\pi f_\pi)^2} \frac{1}{4} \Bigg\{
1-\frac{4-11\tau+3\tau^{2}}{\sqrt{1-\frac{1}{4}\tau}}\sqrt{\tau} \, \arccos\frac{\sqrt{\tau}}{2}
      -6\tau+\tau(-5+3\tau)\ln\tau\Bigg\}  \,,\\
       -\frac{1}{\pi}\int\limits_{-\infty}^0 \dd t^\prime\, \frac{\im \,
  \kappa_n^{(p^3\,\mathrm{loop})}(t^\prime)}{t^\prime - t }&=&
- \frac{g_A^2 M_N^2 }{(4\pi f_\pi)^2} \frac{1}{2} \Bigg\{
2 -\frac{2-\tau}{\sqrt{1-\frac{1}{4}\tau}}\sqrt{\tau} \, \arccos\frac{\sqrt{\tau}}{2}
      -\tau\ln\tau\Bigg\}  \,.
 \eea
 \end{subequations}
The exact same result is obtained by integrating 
 over the Feynman-parameter in the loop expressions of \Eqref{ammloop}. The dispersion
 relation proposed in \Eqref{disprel} is thus verified in this example as well. Note that in this
 case we do not need a subtraction simply because the integral converges. However,
 since the complete result to this order is
 \beq
 \kappa = \, \stackrel{\,\circ}{\kappa} + \kappa^{(p^3\,\mathrm{loop})} + \mbox{c.t.},
 \eeq
 the counter-term contribution, which here is just a constant involving the ``bare" value of a.m.m.,  can be put in correspondence with 
one subtraction at $t=0$.
 
 Let us remark that the same expression for the nucleon a.m.m.\ is obtained as well
 by two other dispersive methods: a derivative of the Gerasimov-Drell-Hearn sum rule
 \cite{Pascalutsa:2004ga}  and a ``sideways dispersion relation" \cite{Holstein:2005db}. 
 Together with the present result, we therefore  already
 have  three {\it different}
 dispersion relations, which can be applied to the a.m.m.\ calculation.
 One can hope that at least one of them will make the two-loop calculation of the
 nucleon a.m.m.\ more
 feasible.

 We conclude the discussion of the nucleon properties with
 the example of the scalar nucleon polarizabilities: $\alpha_N$ (electric) and $\beta_N$ (magnetic). The specifics of 
 this example is that the leading order ($p^3$) correction comes entirely
 from chiral loops, the counter-terms are absent. In the case of magnetic polarizability
 of the proton, the result is given by~\cite{Bernard:1991rq,Lensky:2009uv}:
 \beq
 \beta^{(3)}_p(t) =\frac{e^2 g_A^2}{192\pi^3 f_\pi^2 M_N }
  \int^1_0\! \dd x \, \bigg\{ 1 - \frac{(1-x)(1-3x)^2+x}{(1-x)(t/M_N^2)+x^2 -i0^+} 
  -  \frac{x\, t/M_N^2 +x^2 [1-(1-x)(4-20x+21x^2)]}{\big[(1-x)(t/M_N^2)+x^2 -i0^+\big]^2} 
\bigg\}.
\eqlab{beta}
\eeq
  The imaginary
part can be easily calculated: 
\bea
\mathrm{Im}\, \be^{(3)}_p (t) = - 
\frac{e^2 g_A^2}{192\pi^3 f_\pi^2 M_N }
\frac{\pi }{8\la^3} \Big[ 2-72 \la + (418\la-246) \,\tau - (316\la-471) \,\tau^2 
+(54\la-212)\,\tau^3 +27\tau^4\Big] \,\th(-t),
\eea
and the dispersion relation of \Eqref{disprel} can be shown to hold also for these expressions.
The electric polarizability at order $p^3$ withstands this test too, however the expressions
are more bulky and will be omitted here.

\section{$\Delta$-resonance}
It is interesting to examine a case where the pion production cut
extends into the physical region, as it happens in the case 
when the $\De(1232)$ is included as an explicit degree of freedom
in the chiral Lagrangian (see, e.g., \cite{Jenkins:1991es,Hemmert:1997ye,Pascalutsa:2002pi,Pascalutsa:2006up,Long:2009wq}). In this example the cut may extend from
$t=-\infty$ up to $t=\mathit{\De}^2$, with $\mathit{\De} = M_\De - M_N$, the Delta-nucleon
mass difference. The pion-mass dispersion relation for a static quantity $f$  becomes
\beq
\eqlab{disprel2}
 f(m_\pi^2) = -\frac{1}{\pi}\int\limits_{-\infty}^{\vDe^2} \dd t' \, \frac{\im\, f(t')}{t'-m_\pi^2+i0^{+}}\,.
\eeq

\begin{wrapfigure}{r}{1.9in}
\vspace{-2mm}
\centerline{\epsfclipon   \epsfxsize=4.5cm%
  \epsffile{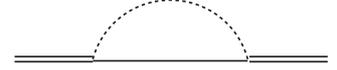} 
}
\caption{A chiral loop correction to the $\De$-isobar mass. Double lines
denote the $\De$ propagation.}
\figlab{deltase}
\end{wrapfigure}
Let us demonstrate how it works on the example of a chiral correction to the 
$\De$-isobar mass. 
A one-loop graph with the cut all the way up to $\vDe^2$ is shown
in \Figref{deltase}. It yields the following
contribution to the self-energy
\cite{Pascalutsa:2005nd}: 
\beq
\Si_\De^{(\pi N\,\mathrm{loop})} (t) = -\frac{1}{2}\left( \frac{h_A M_\De }{8\pi f_\pi} \right)^2   \int_0^1 \dd x\,\, (x M_\De +M_N)\,\mathcal{M}^2(x)\,\Big[ L_\veps + \ln{\mathcal{M}^2(x)}\Big],
\eeq
where $h_A\simeq 2.85 $ is the $\pi N\De$ axial-coupling constant, and
\beq
  \mathcal{M}^2(x) = x^2- (1+r^2-\tau)x +r^2- i 0^+=(x-\al)^2-\la^2-i0^+,
\eeq
with $r = M_N/M_\De$, $\tau = t/M_\De^2$, $\al=\frac{1}{2}(1+r^2-\tau)$, 
and $\la^2= \al^2-r^2$.

The imaginary part arises again from the log, when its argument turns negative in the
region of integration over the Feynman parameter $x$. More specifically,
\beq
\mathcal{M}^2(x) < 0, \quad \mbox{for}\,\, \al-\la < x <  
\left\{ \begin{array}{lc} 1, & \mbox{if}\,\, t<0 \\
\al+\la, & \mbox{if}\,\, 0\leq t \leq \vDe^2\,,
\end{array} \right. 
\eeq
and as the result: 
\beq
\im \, \Si_\De^{(\pi N\,\mathrm{loop})} (t) =
\pi \frac{M_\De}{2}\left( \frac{h_A M_\De }{8\pi f_\pi} \right)^2 \,  \times\,
\left\{ \begin{array}{lc}
\third  (\al+r) \Big[-2 \la^3  + (1-\al)(\tau-2\la^2)\Big] + \frac{1}{4} \tau^2 , & t<0\\
-\frac{4}{3} (\al+r) \la^3, & 0\leq t \leq \vDe^2 \\
0, & t> \vDe^2\,.
\end{array} \right.
\eeq
Note that, despite the awkward separation into regions in $t$, the function
is continuous. For the physical value of the pion mass it provides the
familiar expression for the $\De$-resonance width: $\Gamma_\De = - 2 \,\im \, \Si_\De\simeq 115 $ MeV.

We have checked that, analogously to the nucleon case, this chiral
correction to the $\De$ mass satisfies the doubly-subtracted dispersion relation
in the pion mass squared:
\beq
\eqlab{disprel3}
 \re\, \Si_\De^{(\pi N\,\mathrm{loop})} (m_\pi^2) = 
 -\frac{1}{\pi}\int\limits_{-\infty}^{\vDe^2} \dd t' \, \frac{\im\, \Si_\De^{(\pi N\,\mathrm{loop})} (t')}{t'-m_\pi^2}\left(\frac{m_\pi^2}{t'}\right)^2 \,,
\eeq
where the integration is done, as usual, in the principal-value sense.

\section{Direct calculation of the absorptive part}
Of course to find the absorptive part it should not always be necessary to go through the
entire loop calculation, as we have done so far. In the usual dispersion relations, done in external
variables such as energy, the Cutkowski rules offer a simple method to compute
the absorptive contributions. In our case Cutkowski rules are inapplicable 
because the pion is not on it positive-energy mass shell to begin with. Nevertheless a
direct computation of absorptive parts is possible as will demonstrated in the following three
examples.

Consider first the tadpole, \Figref{nuclse}(b), having the following generic expression
\beq
J_\mathrm{tad}(t) = i\int \frac{\dd^4 k}{(2\pi)^4} \frac{k^{2n}}{k^2-t+i0^+}\,,
\eeq
with an integer $n$.
After the Wick rotation and going to the hyperspherical coordinates we obtain
\beq
J_\mathrm{tad}(t) = \int \frac{\dd \Omega_4}{(2\pi)^4} \int\limits_0^\infty
 \dd K \, \frac{K^{3+2n}}{K^2+t-i0^+}\,,
\eeq
where $ \int \dd \Omega_4 = \int_0^{2\pi} \! \dd\vfi \int_0^{\pi} \! \dd\th  \sin\th  \int_0^{\pi} \! \dd\chi \sin^2\chi
=2\pi^2.$
The absorptive part can now be simply found as
\beq
\im\, J_\mathrm{tad}(t) = \frac{2\pi}{(4\pi)^2} \int\limits_0^\infty
 \dd K \, K^{3+2n} \de(K^2+t) =  \frac{\pi}{(4\pi)^2} (-t)^{1+n} \, \th(-t) \,.
\eeq

As the second example we consider another
typical integral, which appears, e.g., in the calculation of the graph \Figref{nuclse}(a),
\beq
I(t,M^2) =  i\int \frac{\dd^4 k}{(2\pi)^4} \frac{1}{(k^2-t+i0^+) [(p-k)^2-M^2+i0^+]}\,,
\eeq
with $p^2=M^2$.
Again, after the Wick rotation and the adoption of hyperspherical coordinates we obtain
\beq
I(t,M^2) = - \int \frac{\dd \Omega_4}{(2\pi)^4} \int\limits_0^\infty
 \dd K \, \frac{K^2}{(K^2+t-i0^+)(2i M  \cos\chi + K-i0^+)}\,,
\eeq
and therefore the absorptive part is found as 
\bea
\im\, I(t,M^2) &=& - \pi \frac{4\pi}{(2\pi)^4} \int\limits_0^{\pi} \! \dd\chi \sin^2\chi \int\limits_0^\infty
 \dd K \, \frac{K^2}{2i M  \cos\chi + K}\,\de(K^2+t) = 
 - \frac{2}{(4\pi)^2} \int\limits_0^{\pi} \! \dd\chi  \frac{\sqrt{-t}\,\sin^2\chi}{2i M  \cos\chi + \sqrt{-t}}
 \nn\\
 &=& -  \frac{\pi}{(4\pi)^2}\left( \frac{t}{2M^2} +\sqrt{-\frac{t}{M^2} \Big(1-\frac{t}{4M^2}\Big)}\, \right)\,
 \th(-t)\,.
\eea
From these two elementary examples one can see that a direct computation of the absorptive part
is in principle simpler than the one of the full result, since one of the integrals is always lifted by the 
$\de$-function. 

We finally come to a two-loop example, namely the pseudothreshold sunset graph with
two different masses [see \Figref{nuclse}(c)]:
\beq
J_\mathrm{sunset} (m^2,M^2) = \pi^{-d} \! \int\! \dd^d k_1 \int\! \dd^d k_2 \frac{1}{(k_1^2-m^2)(k_2^2-m^2)
[(p-k_1-k_2)^2-M^2]},
\eeq
where $p^2=M^2$. In this case we keep an arbitrary number of dimensions $d=4-2\veps$,
since the absorptive part has an ultraviolet divergence. Conveniently defining
the dimensionless quantities $t=m^2/M^2$ and
\beq
 \tilde J(t) =\frac{M^{2(2\veps-1)}}{\Gamma^2(1+\veps)} J_\mathrm{sunset} (m^2,M^2)\,, 
\eeq
one can show that $\tilde J$ satisfies a hypergeometric type of differential equation~\cite{Argeri:2002wz}:
\beq
t(t-1) \frac{\dd^2 \tilde J(t)}{\dd t^2} + \Big[ \frac{1}{2} - 2\veps +\Big(-\frac{3}{2}+4\veps\Big)\, t\,\Big]\frac{\dd \tilde J(t)}{\dd t}
+ \frac{1}{2}(1-2\veps)(2-3\veps) \tilde J(t) = \frac{1}{2\veps^2}\,\Big( t^{1-2\veps}-2 t^{-\veps}\Big).
\eeq 
As the boundary conditions one may use the easily computable massless expressions:
\bea
J_\mathrm{sunset} (m^2,M^2=0) &=& -m^{2(1-2\veps)} \frac{\Gamma^2(1+\veps)}{\veps^2 (1-\veps)(1-2\veps)}\,, \\
J_\mathrm{sunset} (m^2=0,M^2) &=& M^{2(1-2\veps)}\, \frac{\Gamma(3-4\veps)\,\Gamma(2\veps-1) \,\Gamma^2(1-\veps)\,\Gamma(\veps)}{ \Gamma(2-2\veps) \,\Gamma(3-3\veps)}\,,
\eea
hence, e.g.,  $t^{-1+2\veps} \tilde J(t) \stackrel{t\to+\infty}{=} -1/[\veps^2 (1-\veps)(1-2\veps)]$.

Since for real $t$ the equation is linear with real coefficients we deduce that the solution develops an imaginary part when the inhomogeneous term (the r.h.s.) develops an imaginary part, i.e.,
for $t<0$. The existence of the cut for negative $m^2$ is evident. 
Furthermore, when seeking the solution in the $\veps$-expanded form, we again observe that
solving for the imaginary part is simpler, because the corresponding inhomogeneous term
is simpler at any given order in $\veps$, cf.: 
\bea
\frac{1}{2\veps^2} t^{-\veps}(t^{1-\veps}-2)
&=& \frac{\half t-1}{\veps^2} -\frac{t-1}{\veps} \ln t + \frac{2t-1}{2} \ln^2 t 
+ \cO(\veps),\\
 \im\, \frac{1}{2\veps^2} t^{-\veps}(t^{1-\veps}-2) & = & \th(-t) \, \pi\, \Big[
\frac{ t-1}{\veps}  - (2t-1) \ln(- t) 
+ \cO(\veps) \Big].
\eea

The solution for the imaginary part is of the form
\beq
\im\, \tilde J(t) = \th(-t) \, \pi\, \Big[ -\frac{2t}{\veps} + t\, \Big(-7 +(2+t) \ln(-t)\Big) - (1-t)^2 \ln(1-t) + \cO(\veps)\Big],
\eeq
which agrees with the result derived from a conventional formulae \cite{Berends:1997vk}.

\section{Concluding remarks}
Hadron properties depend  on the pion-mass squared (or, the light-quark mass, $m_{u,d}\sim m_\pi^2$) in an essentially non-analytic way. 
In this work we have identified the origin of this non-analyticity with a cut
in the $m_\pi^2$ complex plane, which extends along the negative axis. In $\chi$PT, the cut arises
due to the possibility of a ``subsoft'' pion production.
Assuming analyticity in the rest of the complex plane, we are able to
write a simple dispersion relation in $m_\pi^2$, cf.\ \Eqref{disprel}.
The validity of this relation has been tested here, on a number of quantities
computed in $\chi$PT to lowest order. It also has been tested here on a generic two-loop
example.

There are at least two ways in which the proposed dispersion relation can be useful.
First, as a consistency constraint of various ``chiral extrapolation'' formulas  and methods.
Second, as a computational technique, similarly to how the usual dispersion relations, written in terms of energy (or,
in relativistic theory, the Mandelstam variables), are used. 

Although, the usual dispersion relation appear to be quite different
from the dispersion relation in the mass, at a given kinematical point, the Mandelstam variables can take values given entirely in terms of mass, e.g., $t=m_\pi^2$. Then, 
the dispersion relation in that variable can be used to connect to another kinematical
point, e.g., $t=0$. This is precisely the strategy that had long ago been proposed
to relate the scattering amplitudes at the physical pion mass with their 
chiral-limit values \cite{Fubini:1968,deAlfaro:1968}.
It led to the so-called ``mass-dispersion
 relations" \cite{Gordon:1970dn,Murtaza:1970xe,O'Donnell:1972mc}, 
 which appear to be similar to the relation put forward in this work.\footnote{Comparing Eq. (2.1) in Ref.~\cite{Fubini:1968} with \Eqref{disprel} here, one can see that the generic form is the same.} Whether the similarities extend beyond the general form is not easy to tell since the analytic properties of the pion-mass dependence appear to be much more involved 
in the case of the old mass-dispersion relations.
 Independently of whether there is a connection between the old and new 
 pion-mass dispersion relations, 
their test and validation in $\chi$PT has been made only now.
 
\section*{Acknowledgments}
We thank  Akaki Rusetsky 
 for valuable remarks on the manuscript.  
This work is partially supported by 
Deutsche Forschungsgemeinschaft (DFG).
The work of T.~L.\ is partially supported by the Research Centre
"Elementarkr\"afte und Mathematische Grundlagen" at the Johannes Gutenberg
University Mainz.

\end{document}